\documentclass[10pt,letterpaper,superscriptaddress]{article}
\usepackage{opex3}
\usepackage{amsmath}
\usepackage{amsthm}
\usepackage{cite}

\newcommand{\ket}[1]{\left\vert#1\right\rangle}
\newcommand{\bra}[1]{\langle #1\vert}

\begin{document}
\vskip4pc

\clearpage

\title{Experimental quantum teleportation over a high-loss free-space channel}

\author
{Xiao-song Ma,$^{1,2*}$ Sebastian Kropatschek,$^{1}$ William Naylor,$^{1}$ Thomas Scheidl,$^{1}$ Johannes Kofler,$^{1,4}$ Thomas Herbst,$^{3}$ Anton Zeilinger,$^{1,2,3}$ Rupert Ursin,$^{1**}$\\}

\address{$^{1}$Institute for Quantum Optics and Quantum Information (IQOQI), Austrian Academy of Sciences, Boltzmanngasse 3, A-1090 Vienna, Austria, \\
              $^{2}$Vienna Center for Quantum Science and Technology, Faculty of Physics, University of Vienna, Boltzmanngasse 5, A-1090 Vienna, Austria,\\
              $^{3}$Faculty of Physics, University of Vienna, Boltzmanngasse 5, A-1090 Vienna, Austria,\\
              $^{4}$Present address: Max Planck Institute of Quantum Optics, Hans-Kopfermann-Str. 1, 85748 Garching/Munich, Germany.}
\email{$^{*}$Xiaosong.Ma@univie.ac.at}
\email{$^{**}$Rupert.Ursin@univie.ac.at}

\date{\today}

\begin{abstract}
We present a high-fidelity quantum teleportation experiment over a high-loss free-space channel between two laboratories. We teleported six states of three mutually unbiased bases and obtained an average state fidelity of $0.82(1)$, well beyond the classical limit of 2/3. With the obtained data, we tomographically reconstructed the process matrices of quantum teleportation. The free-space channel attenuation of 31~dB corresponds to the estimated attenuation regime for a down-link from a low-earth-orbit satellite to a ground station. We also discussed various important technical issues for future experiments, including the dark counts of single-photon detectors, coincidence-window width etc. Our experiment tested the limit of performing quantum teleportation with state-of-the-art resources. It is an important step towards future satellite-based quantum teleportation and paves the way for establishing a worldwide quantum communication network.
\end{abstract}


\section{Introduction}

Teleportation of quantum states~\cite{Bennett1993,Bouwmeester1997} and its remote state preparation variant~\cite{Boshi1998} are intriguing
concepts within quantum physics and striking applications of
quantum entanglement. Introduced as ``a term from science fiction,
meaning to make a person or object disappear while an exact
replica appears somewhere else''~\cite{Bennett1993}, quantum teleportation has now
become a crucial building block for many quantum information
processing schemes. Besides their importance for quantum
computation~\cite{Gottesmann1999,Knill2001,Ladd2010}, teleportation and entanglement swapping~\cite{Yurke1992,Zukowski1993} are at the
heart of the quantum repeater~\cite{Briegel1998,Duan2001}, allowing to
distribute quantum entanglement over long distances, thus being
vital for global quantum communication schemes. Quantum
teleportation over longer distances will be needed for the realization of
quantum network schemes among several parties~\cite{Bose1998}, which connect devices utilizing quantum computational algorithms.

To realize global quantum communications, qubits must be transferred over long
distances. Most earlier teleportation experiments are in-lab demonstrations, and hence the communication distance is rather limited. Although fibre-based, long-distance teleportation
has been studied experimentally~\cite{Marcikic2003,Ursin2004}, the maximum transmission distance will be ultimately limited
by intrinsic photon losses in optical fibre. On the other hand, free-space
channels are ideal for transmitting photonic qubits due to the low absorption and the weak birefringence of the atmosphere. Moreover, in outer space, photon loss and decoherence are
negligible. Therefore, by using satellites, optical free-space links potentially allow much larger photon propagation distances.

In recent years, there has been significant
progress in developing free-space optical links for applications in quantum communications~\cite{Hughes2002,Kurtsiefer2002,Aspelmeyer2003,Resch2005,Peng2005,Schmitt-Manderbach2007,Ursin2007,Villoresi2008,Fedrizzi2009,Scheidl2009,Jin2010,Scheidl2010}. However, all these previous experiments are utilizing either a single photon or a pair of entangled photons. A multiphoton, free-space and long-distance quantum teleportation remains an experimental and technical challenge. This is because of the attenuation of the link, which drastically reduces the detection rate associated with the simultaneous detection of four photons and hence decreases the signal-to-noise ratio.

Here we present an experimental demonstration of quantum teleportation over a high-loss free-space channel, where the two involved
laboratories, Alice and Bob, are separated by 10~m in the basement of one of the buildings owned by Austrian Academy of Sciences (see Fig. 1). We simulated the link attenuation by inserting neutral density filters and obtained up to 36~dB attenuation. Even under such a high-loss condition, we successfully achieved quantum teleportation.

\section{Quantum teleportation}

The task of quantum teleportation is the following. Alice wants to send an unknown quantum state of a particle, provided by Charlie, to Bob. A trivial way to accomplish that would be to send the particle itself. If transferring the original particle itself is not possible, she can attempt to clone the quantum state of this particle onto another particle and send it to Bob. Due to the no-cloning theorem~\cite{Wootters1982}, this procedure will not work with 100\% efficiency. However, quantum teleportation, which is based on both a quantum channel and a classical channel shared by Alice and Bob, circumvents the no-cloning theorem and provides a solution for such a task. A detailed explanation can be found in~\cite{Bennett1993,Bouwmeester1997}.

In our case, an entangled photon pair is distributed over the quantum channel. We use the polarization-entangled state
\begin{equation} \label{bellstate}
|\Psi^- \rangle_{12}  =
\frac{1}{\sqrt{2}}(|H\rangle_1 |V\rangle_2 - |V\rangle_1
|H\rangle_2),
\end{equation}
which is one of the four maximally entangled Bell states.
$|H\rangle_j$ and $ |V\rangle_j$ denote the horizontal and
vertical polarization states of photon $j$. Alice and
Bob share this entangled state (shown in Eq.(~\ref{bellstate})), in that photon 1 is with Alice
and photon 2 is with Bob. In order to transfer the
unknown polarization state of Charlie's photon 3
\begin{equation} \label{photon3}
|\phi\rangle_3=\alpha
\ket{H} + \beta \ket{V},
\end{equation}
Alice performs a joint Bell-state measurement (BSM) on photons 1 and 3.
She will randomly observe one out of the four Bell-states
($\ket{\Psi^\pm}_{13}=\frac{1}{\sqrt{2}}(|H\rangle_1 |V\rangle_3 \pm |V\rangle_1
|H\rangle_3)$ and $\ket{\Phi^\pm}_{13}=\frac{1}{\sqrt{2}}(|H\rangle_1 |H\rangle_3 \pm |V\rangle_1|V\rangle_3)$), each with the same
probability of $25\%$. Then she employs the classical channel and conveys the results of the BSM to Bob. With linear optics we can only identify
two out of four Bell states unambiguously~\cite{Calsamiglia2001}. In our case, we choose to identify $\ket{\Psi^-}$ and $\ket{\Psi^+}$, and neglect the results of $\ket{\Phi^+} $ and $\ket{\Phi^-}$ since we can't identify them with certainty. Note that in the other 50\% of the cases, our BSM results in a projection onto a product state (i.e. $|HH\rangle$ or $|VV\rangle$), and hence quantum teleportation is impossible.

If Alice detects $\ket{\Psi^-}_{13}$, the state
of Bob's photon (photon $2$) will be the same as the initial state of photon $3$, as shown in Eq.~(\ref{photon3}), except for an irrelevant
global phase factor. Therefore, there will be nothing left to do
for Bob to complete the teleportation protocol. In the
$\ket{\Psi^+}_{13}$ case, the conditional state of Bob's photon is $\alpha \ket{H} -
\beta \ket{V}$ and thus he has to apply a relative $\pi$ phase
shift between the polarization components $\ket{H}$ and $\ket{V}$
to convert the state of photon $2$ into the original state of
photon 3. Hence, in order to complete quantum teleportation, Alice must communicate via classical means to Bob which state she observed
in her BSM. In accordance to the result of the BSM, Bob can apply the
unitary transformation on photon $2$ and consequently obtain an exact replica
of the quantum state of the input photon $3$.

\section{Experiment and results}
\begin{figure}[h!]
\centerline{\includegraphics[width=1\textwidth]{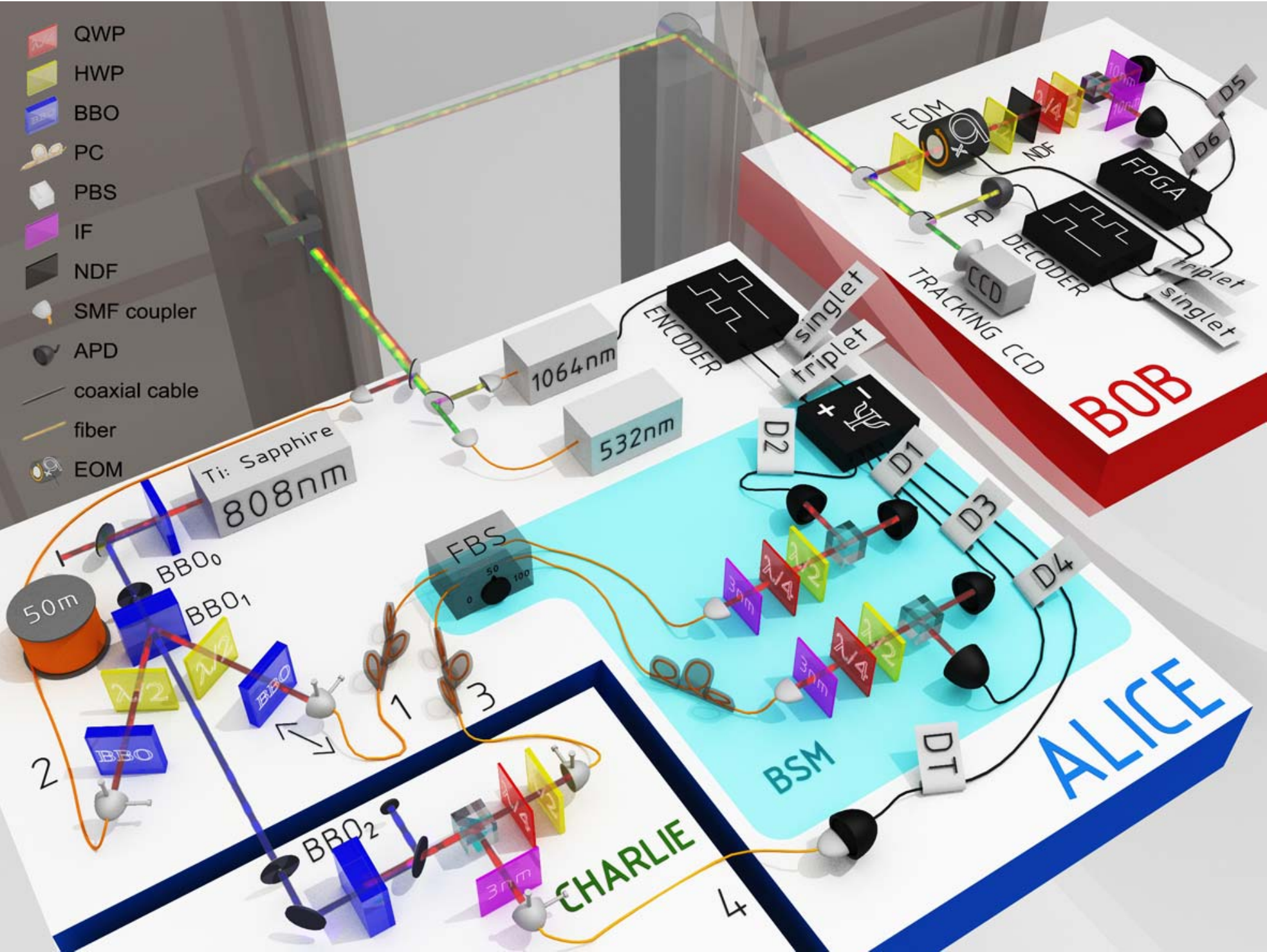}}\caption{The setup for quantum teleportation over a high-loss channel. We up-convert near-infrared femtosecond pulses (central wavelength of 808~nm) emitted from a mode-locked Ti:Sapphire laser to blue pulses (central wavelength of 404~nm) via a $\beta$-barium borate crystal (BBO$_0$). Polarization-entangled photon pairs, photons 1 and 2, are produced by using photon emissions of a non-collinear type-II spontaneous parametric down conversion from BBO$_1$ and sent to Alice and Bob. Photon 2 is delayed with a 50~m single-mode fiber and then sent to Bob. The input photon of teleportation, photon 3, is generated from BBO$_2$ via a collinear heralded single-photon source. Alice performs a Bell-state measurement (BSM) on photons 1 and 3 via fiber beam splitter (FBS) and can unambiguously identify the results of singlet ($\ket{\Psi^-}_{13}$) or triplet ($\ket{\Psi^+}_{13}$). These BSM results are then encoded in a 1064~nm laser via an encoder and sent to Bob. We use a 532~nm laser and a charge-coupled device (CCD) camera to simulate the tracking and pointing system, which will be required in an experiment between satellites and ground stations. The beams of photon 2, of the 1064~nm laser and of the 532~nm laser are multiplexed to a common optical path by means of various dichroic mirrors at Alice's side and propagate along the hallway connecting two separate labs and through keyholes of of both doors. On Bob's side, we demultiplex the three beams and perform feed-forward operations with a photodetector (PD), a decoder and an electro-optical modulator (EOM). The 3~nm interference filters (IF) are employed to eliminate the spectral distinguishability of single photons. The combination of a quarter-wave plate (QWP), a half-wave plate (HWP), a polarization beam splitter (PBS) and avalanche photodiodes (APD, D1-D6) is used to measure the polarization state of single photons. Various neutral density filters (NDF) are used to vary the attenuation of the link. Polarization controllers (PC) are used to compensate the unwanted polarization rotation induced by fibers. See text for details.} \label{Setup}
\end{figure}
The setup for the experimental demonstration of quantum teleportation over a high-loss free-space channel is shown in Fig.\ \ref{Setup}. The pump laser is a mode-locked Ti:sapphire femto-second laser with a pulse duration of 140~fs and a repetition rate of 80 MHz. The central wavelength of the pump beam is at 808 nm. A $\beta$-barium borate crystal (BBO$_0$) is used to up-convert the pump pulses to blue pulses via second harmonic generation. The up-converted blue pulses' central wavelength is 404 nm. We use several dichroic mirrors to separate the blue pulses from the remaining infrared pulses.

Photons 1 and 2 are generated from BBO$_1$ via spontaneous parametric down conversion (SPDC) in a non-collinear type-II phase matching configuration~\cite{Kwiat1995}. After walk-off compensation with half-wave plates (HWP) and compensation BBO crystals, they are adjusted to be the polarization-entangled state, $|\Psi^- \rangle_{12}$. Photons 3 and 4 are generated from BBO$_2$ in a collinear type-II phase matching configuration and are separated by a polarizing beam splitter (PBS). Charlie prepares the quantum state of photon 3, which is the original teleportee. Photon 2 is sent to Bob and will become the final teleported one. Without attenuation, the local detected count rates of the two photon pairs were approximately 90 and 110 kHz for the entanglement resource and the single photon input respectively.

All the photons are coupled into single-mode fibers. We guide photons 1 and 3 to a fiber-based beam splitter, where two-photon non-classical interference occurs. Also using polarization-resolving photon detections, we can perform a BSM which projects photons 1 and 3 randomly onto either $\ket{\Psi^-}_{13}$ or $\ket{\Psi^+}_{13}$. In order to obtain a high-quality result of the BSM, we eliminate the temporal, polarization and spectral distinguishabilities. The relative temporal delay between photons 1 and 3 is adjusted with a motorized translation stage mounted on the fiber coupler of photon 1. Fiber polarization controllers (PC) are employed to eliminate the polarization distinguishability of the two interfering photons. Interference filters (full width at half maximum of about 3 nm) are used to eliminate the spectral distinguishability. All the photons are detected by silicon avalanche photodiodes (APD).

The $|\Psi^-\rangle_{13}$ state can be identified with coincidences between detectors D1 and D4 or D2 and D3, because photons 1 and 3 leave the beam splitter in different outputs and exhibit orthogonal polarization. In the $|\Psi^+\rangle_{13}$ case, the photons will take the same output of the beam splitter, however they will still have orthogonal polarization in the $\ket{H}/\ket{V}$ basis. Therefore, we can identify $|\Psi^+\rangle_{13}$ by coincidences between detectors D1 and D2 or D3 and D4.

When Alice obtained a $\ket{\Psi^-}$, she encodes this result with a strong single-light-pulse signal with an encoder and sends this classical information to Bob. On the other hand, when she obtained a $\ket{\Psi^+}$, she encodes it with a double-light-pulse signal and sends that over to Bob. The classical light pulses are generated from a 1064 nm laser, which is modulated by the encoder. Then we spatially combine the optical paths of photon 2, the 1064 nm and a 532 nm laser beam (simulating a tracking beacon laser) on two dichroic mirrors. Photon 2 is delayed with a 50 m single-mode fiber, which gives Bob enough time for performing the local unitary transformation (i.e. a relative $\pi$ phase shift between the polarization components $\ket{H}$ and $\ket{V}$) if Alice obtains $\ket{\Psi^+}$ as the BSM's result.

We link Alice's and Bob's labs via a free-space channel. The optical path of the beam is through the keyholes of both doors of two separate labs and along the hallway of the basement of one of the buildings owned by Austrian Academy of Sciences. In Bob's lab, we separate all three signals, the single photons (808 nm), the 1064 nm and the 532 nm beams using two dichroic mirrors. Furthermore, the 1064 nm beam detected with a photodetector (PD) and the encoded results of the BSM are then extracted by Bob with a decoder. The 532 nm beam is monitored by a charge-coupled device (CCD) camera, which would be used for the tracking and pointing mechanism in the experiments between satellites and ground stations.

As already mentioned, Bob will directly obtain the original state
on photon $2$, if Alice observes the $|\Psi^- \rangle_{13} $
state. And should Alice observe the $|\Psi^+ \rangle_{13} $ state, Bob will have to introduce a phase shift of $\pi$ between
$|H\rangle_2$ and $ |V\rangle_2$ polarization of photon $2$.  In
our experiment, this transformation is achieved with a fast
electro-optical modulator (EOM). When Alice's BSM result was
$|\Psi^- \rangle_{13} $, then Bob left the EOM at the idle voltage (the EOM
introduces no phase shift), and if it was a $|\Psi^+ \rangle_{13} $,  then Bob
applied the half-wave voltage (the EOM introduced the relative $\pi$
phase shift between the $\ket{H}$ and $\ket{V}$ polarization
component) just before the photon passed through the EOM.

\begin{figure}[h!]
\centerline{\includegraphics[width=0.85\textwidth]{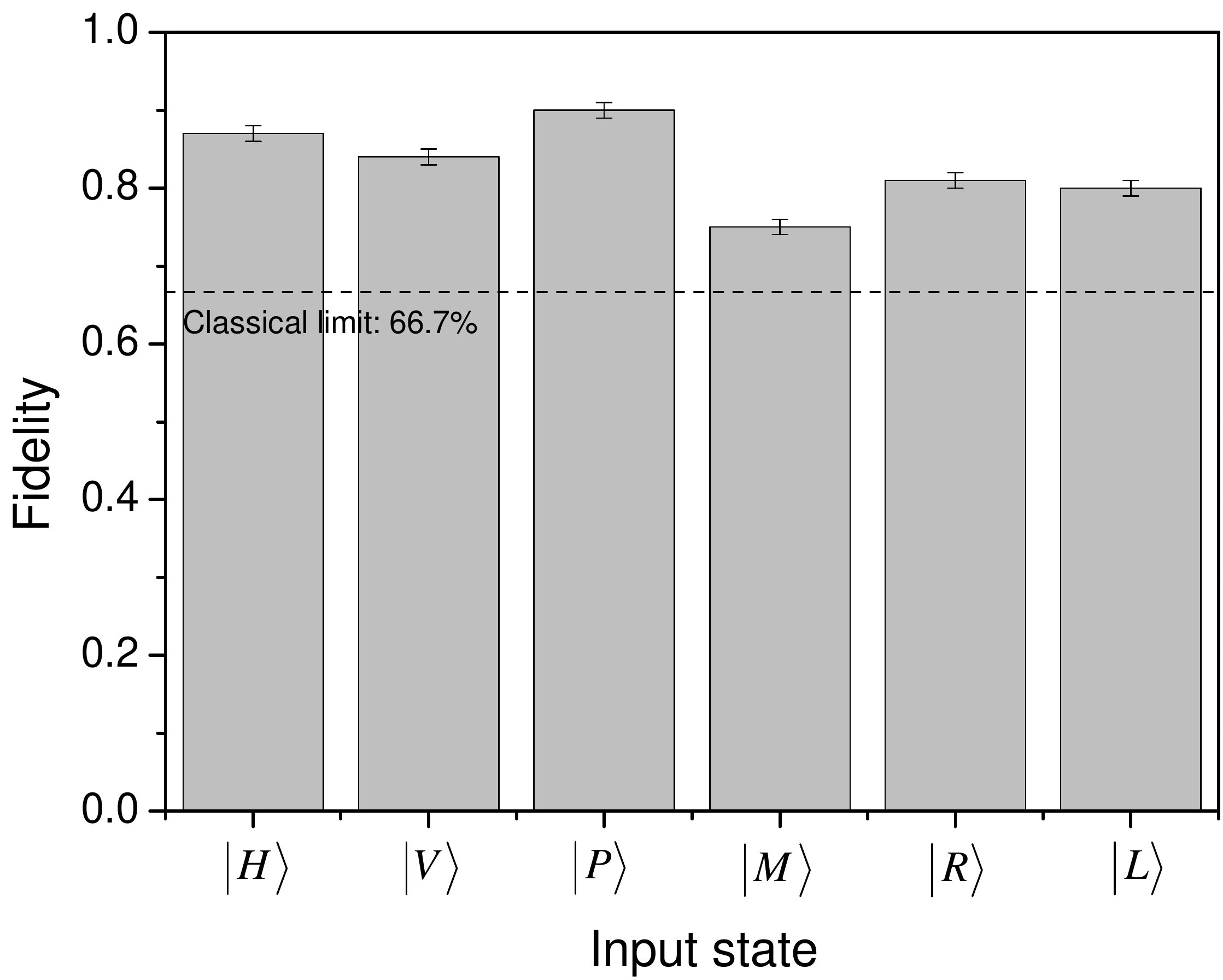}}\caption{State fidelity results for the six unbiased-basis states teleported from Alice to Bob over a high-loss (-31 dB) free-space channel. The observed fidelities, $f$, of the teleported quantum states are: $\ket{H}$ with fidelity $f=0.87 (1)$, $\ket{V}$ with $f=0.83(1)$, $\ket{P}=\ket{H}+\ket{V}$ with $f=0.90 (1)$, $\ket{P}=\ket{H}-\ket{V}$ with $f=0.74 (1)$, $\ket{R}=\ket{H}+i\ket{V}$ with $f=0.81 (1)$, $\ket{L}=\ket{H}-i\ket{V}$ with $f=0.80 (1)$. All fidelities significantly exceed the average classical limit of 2/3. The data shown comprise a total of 9891 four-fold coincidence counts in about 50 hours summed over all input states. Error bars are given by Poissonian statistics.} \label{StateFed}
\end{figure}

Figure\ \ref{StateFed} shows the state fidelity results of quantum teleportation over a free-space channel with $-31$ dB attenuation. We teleported and performed tomographical measurements on the
set of six mutually unbiased bases states $\ket{\phi_{ideal}}\in$
\{ $\ket{H}$, $\ket{V}$, $\ket{P}=(\ket{H}+\ket{V})/\sqrt{2}$, $\ket{M}=(\ket{H}-\ket{V})/\sqrt{2}$, $\ket{R}=(\ket{H}+i\ket{V})/\sqrt{2}$, $\ket{L}=(\ket{H}-i\ket{V})/\sqrt{2}$ \} at the Bob's side (photon 2).
The density matrix, $\rho$, for each of
these teleported states is reconstructed from the experimentally obtained data using the maximum-likelihood
technique~\cite{White1999, James2001}. The
fidelity of the teleportation is defined as the overlap
of the ideal teleported state and the measured
density matrix $f=\left\langle\phi_{ideal}\vert\rho\vert\phi_{ideal}\right\rangle$. For this set
of states the teleported state fidelities are measured to be $f = {0.87(1), 0.83(1),
0.90(1), 0.74(1), 0.81(1), 0.80(1)}$, yielding an
average teleportation fidelity $\bar{f} =0.82(1)$. The
average classical fidelity limit of 2/3~\cite{Bennett1993,Popescu1994} is clearly
surpassed by our observed fidelities. Therefore, we explicitly demonstrated quantum teleportation.

\begin{figure}[h!]
\centerline{\includegraphics[width=0.8\textwidth]{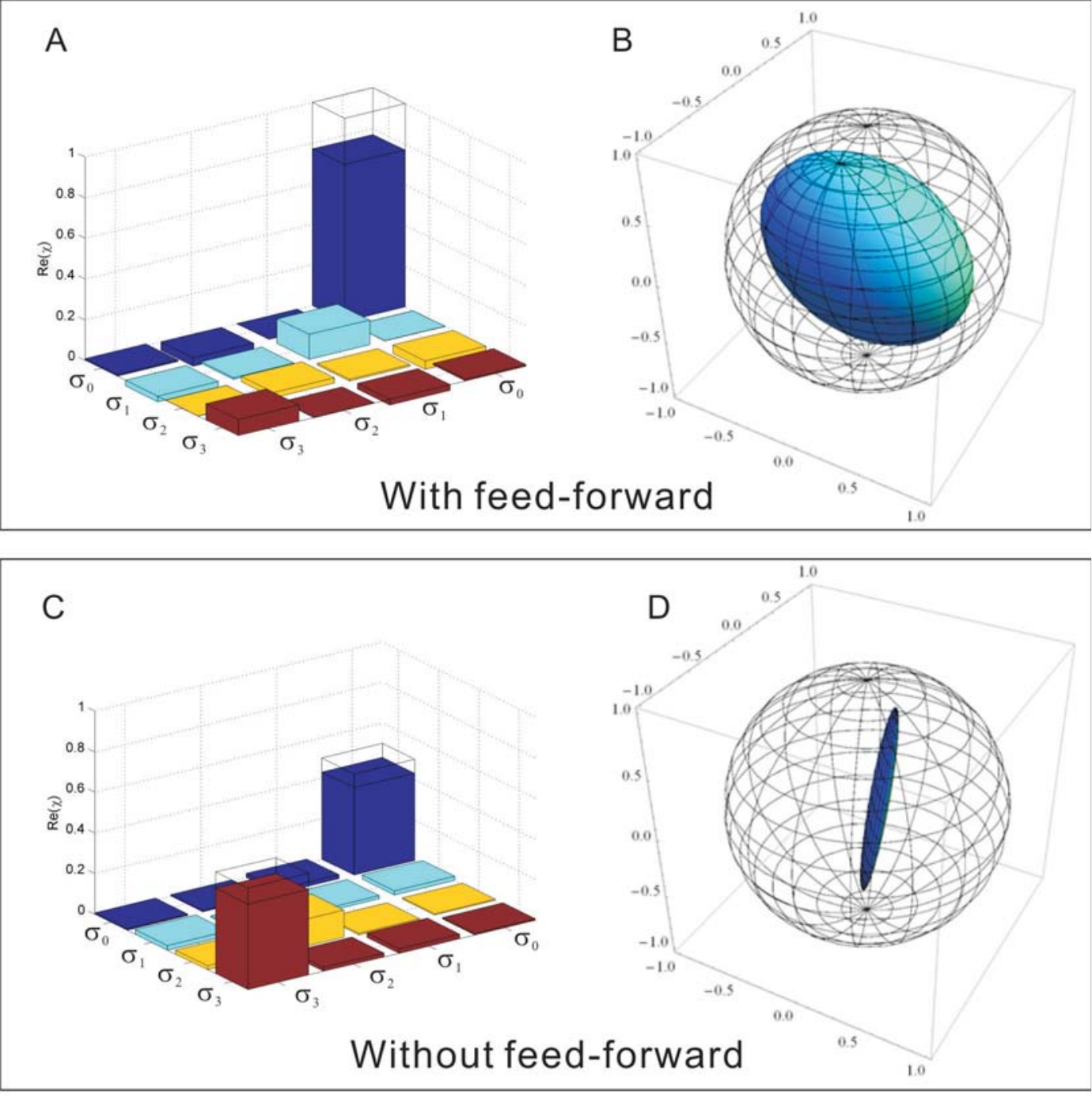}}\caption{(A) The real part of the reconstructed quantum process matrix, $Re(\chi_{lk})$, with $l$, $k$ = 0, 1, 2, and 3 with feed-forward operation. The process matrix of quantum state teleportation is reconstructed from the state tomography of the six mutually unbiased bases states teleported between Alice and Bob. The operators $\sigma_{i}$ are the identity (\textit{i} = 0) and the \textit{x}-, \textit{y}-, and \textit{z}-Pauli matrices (\textit{i} = 1, 2, 3). As intended, the dominant component of $\chi_{lk}$ is the contribution of the identity operation, yielding an overall process fidelity $f_{process} =tr(\chi_{ideal}\chi) = 0.77(1)$. The plot in (B) shows how the input states
lying on the surface of the initial Bloch sphere (meshed surface) are transformed
by our teleportation protocol, with the output states lying on the solid blue surface. (C) The real part of the reconstructed quantum process matrix, $Re(\chi_{lk})$, with $l$, $k$ = 0, 1, 2, 3 without feed-forward operation. The quantum process fidelity $f_{process} =tr(\chi_{ideal}\chi) = 0.24(2)$. The resultant low fidelity is due to the lack of feed-forward operation. This can be also visualized in (D), where the pure input states (meshed surface) are transformed into a mixture (solid blue surface).} \label{Processtomo}
\end{figure}

Moreover, the reconstructed density matrices of the teleported quantum states over the free-space channel with $-31$~dB attenuation allow us to fully characterize the teleportation procedure by quantum process tomography. The four input states ($\ket{H}, \ket{V}, \ket{P}, \ket{R}$) and their corresponding (reconstructed) output states are used to compute analytically the process matrix of quantum teleportation~\cite{NielsenChuang2000}. We can completely describe the effect of teleportation on an input state $\rho_{in}$ by determining the process matrix $\chi$, defined by $\rho = \sum_{l,k = 0}^3 \chi_{lk} \hat{\sigma}_l \rho_{in} \hat{\sigma}_k$, where the $\sigma_{i}$ are the Pauli matrices with $\sigma_{0}$ the identity operator, respectively. To evaluate our process, we take $\rho_{in} = \ket{\phi_{ideal}} \bra{\phi_{ideal}}$. The ideal process matrix of quantum teleportation, $\chi_{ideal}$, has only one nonzero component, $\left( \chi_{ideal} \right)_{00} = 1$, meaning the input state is teleported without any reduction in fidelity. Fig.~\ref{Processtomo}(A) shows the real part of $\chi$ for quantum teleportation with the feed-forward operation. The process fidelity is $f_{process} = \text{tr} \left( \chi_{ideal} \chi \right) = 0.77(1)$.

A quantum process operating on a single qubit can be conveniently represented graphically by visualizing the deformation of a Bloch sphere subjected to the quantum process~\cite{NielsenChuang2000}. As shown in Fig.~\ref{Processtomo}(B), we represent the initial input states as the states lying on the meshed surface of the Bloch sphere. After teleportation, the initial Bloch sphere is deformed into an anisotropic ellipsoid with reduced radius, which corresponds to a reduction of the purity of the teleported state. The reduction of the purity is mainly due to the mixture coming from higher-order emission in the down-conversion process. Note that one can eliminate the degradation of the quantum teleportation process stemming from higher-order emissions from the source of photon 3 by tomographically measuring the input state. And then one can reconstruct the process using the density matrices of the input and output states, instead of using the ideal and output states.

The classical channel and the feed-forward operation are of crucial importance. Without them, Bob cannot take into account the result of Alice's BSM and thus his quantum state is completely mixed in two of the three mutually unbiased bases and hence the classical limits are not surpassed. In this case, quantum teleportation is not successful. For demonstrating this, we obtained data without feed-forward operation. In this case, Bob observes only a
completely mixed output state for the input states $\ket{P}$, $\ket{M}$, $\ket{R}$ and $\ket{L}$ (and all other states in the equatorial plane). However for the $\ket{H}$ or
$\ket{V}$ polarized state, the lacking feed-forward is irrelevant. Thus, the teleportation of any $\ket{H}$ or $\ket{V}$ state (i.e. the north and south pole of the Bloch sphere) will remain unaffected by the lack of feed-forward. The quantum process matrix of teleportation without feed-forward operation is shown in Fig.~\ref{Processtomo}(C). The main components of the process matrix without feed-forward are the identity ($\sigma_{0}$) and the phase flipping ($\sigma_{3}$) operation. The identity operation component is the result of $\ket{\Psi^-}$ as the outcome of the BSM, while the phase flipping operation is the result of $\ket{\Psi^+}$ as the outcome of the BSM. As a consequence of the equal mixture of these two operations, the initial input states (the meshed surface in Fig.~\ref{Processtomo}(D)) are transformed to output states forming a cigar-shaped ellipsoid near the axis of the Bloch sphere (solid blue surface). In the equatorial plane of the sphere, the radii are close to zero showing that the output states are mixed in the $\ket{P}/\ket{M}$ and the $\ket{R}/\ket{L}$ bases. In the $\ket{H}/\ket{V}$ basis, it is unaffected as the radii along the axis pointing to both poles are significant. Note that both output state distributions---the cigar-shape ellipsoid in the case of no feed-forward as well as the more spherical distribution in the case of feed-forward---are tilted with respect to the axis of the Bloch sphere of the ideal initial states. These phenomena are mainly due to polarization drifts of photon 2 in the 50 m delay fiber over long integration time.

\section{Discussion and conclusion}

Quantum teleportation over a high-loss channel is experimentally challenging, because the detection rate is very low, stemming from (1) the intrinsically low rate of simultaneous emission of four photons from the source, and (2) the attenuation of the link. Note that by using periodically-poled crystals, continuous-wave laser pumped SPDC sources could be brighter in the two-fold count rate~\cite{Fedrizzi2009,Scheidl2010}. But in order to perform multi-photon experiments based on CW pumped system, the timing jitter of commercially available Si-APD (tens to hundreds of ps) requires very narrow filtering of the down-converted single photons~\cite{Halder2007}, which drastically reduces the useful four-fold coincidence counts and hence makes the realization of quantum teleportation over a high-loss channel even harder. The two main challenges in our experimental realizations of teleportation over a high-loss link are:
\begin{enumerate}
  \item the extremely low signal-to-noise ratio (SNR) at Bob;
  \item the long integration times needed due to the reduced signal.
\end{enumerate}

In order understand the influence of the attenuation to the teleportation, we developed an analytical model to derive the expected quality based on the photon statistics of SPDC. This model includes the main background contributions for a four-fold coincidence count: second-order emission from SPDC and dark counts from Bob's detectors, and third-order emission. The probability of successful teleportation is given as the product of a successful Bell state measurement at Alice, $p_{\textrm{BSM}}$ and the link efficiency from Alice to Bob, $\eta$. The corresponding probability to record an error due to noise at Bob is $p_{\textrm{BSM}}\cdot n \cdot \tau$, where $n$ is the dark count rate at Bob and $\tau$ is the coincidence-window width in time. Thus the SNR is given by

\begin{equation}
	\textrm{SNR} = \frac{\eta}{n}\cdot\frac{1}{\tau} \, .
	\label{eqSNR}
\end{equation}

\noindent

\begin{figure}[htbp]
	\centering
	\hspace{0in}
 	\includegraphics[width=1\textwidth]{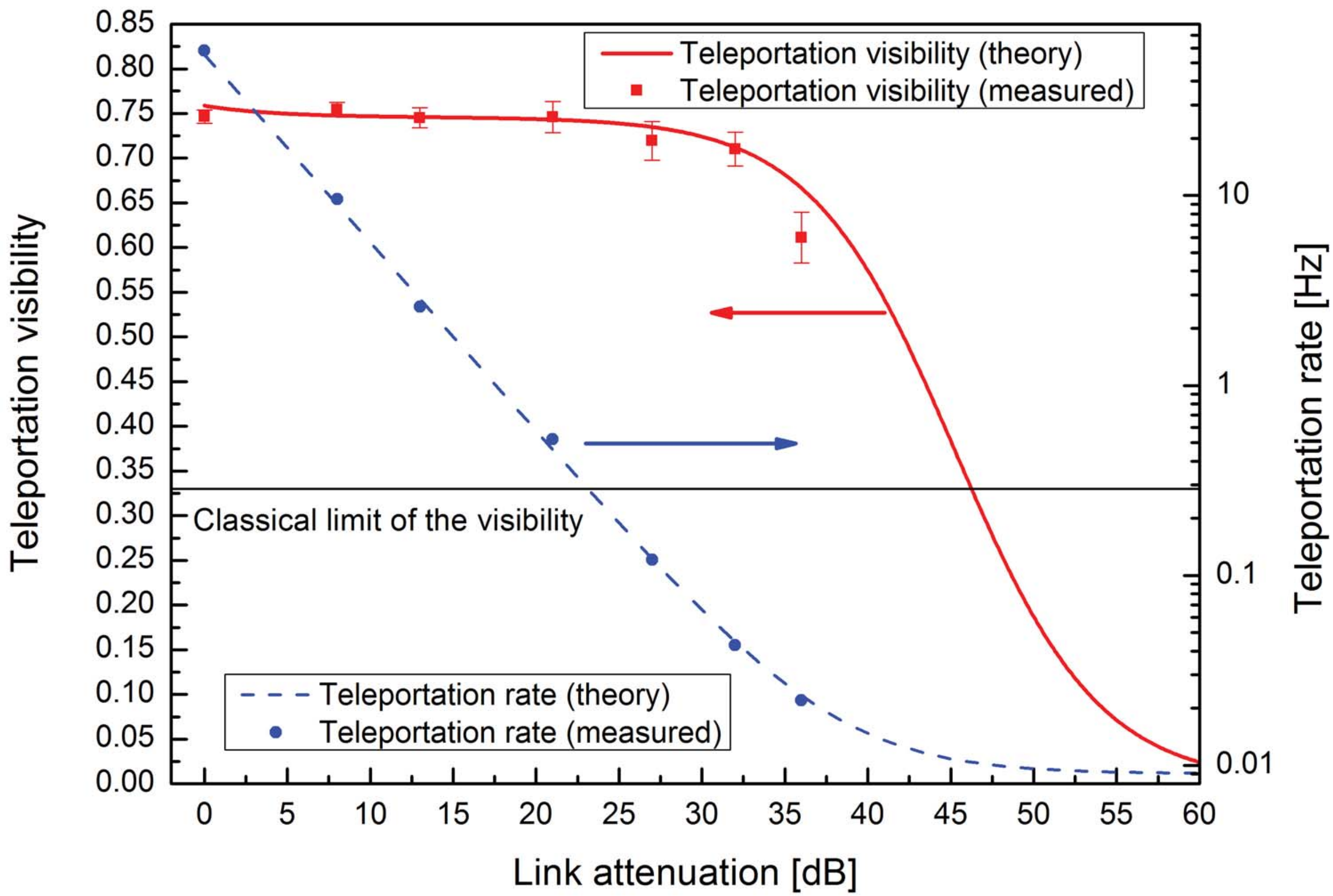}
	\caption{Teleportation visibility and count rate depending on the link attenuation. The measurements were completed on a setup in a single room and without feed-forward, but otherwise the same as the final experiment. The red squares and the blue circles are the experimentally obtained visibilities and teleportation rate (4-fold coincidence count rate), respectively. The red and blue curves are the predictions of our model. Note that the teleportation rate becomes steady above 50 dB. This is because the dominant contribution to the teleportation rate in this attenuation regime is a four-fold coincidence arising from a three-fold coincidence at Alice's side and a dark count at Bob's side. The error bars are calculated based on a Poissonian distribution.}
	\label{figHighDB}
\end{figure}

In Fig.\ \ref{figHighDB}, we show the teleportation visibility in the case where Charlie's input state always was $\ket{P}$. (Based on the results obtained above, we assume that the quality of our teleportation device is the same for all bases.) We picked the state $\ket{P}$, because it is a coherent superposition of the eigenstates of the BSM, which is experimentally challenging, as it requires non-classical two-photon interference. If, in a classical teleportation protocol, Alice does not know the input state, she has to randomly pick a measurement basis and send a corresponding photon to Bob. Then the classical average teleported stated fidelity will be $\bar{f}_{cl} = 2/3$. The classical average visibility is given by $\bar{V}_{cl} = 2\bar{f}_{cl}-1 = 1/3$. Therefore, if we experimentally obtained a visibility above 1/3, we achieved quantum teleportation.

One can see that with our current parameters (the achievable 4-fold count rate, APD dark count rates, coincidence-window width etc), teleportation is feasible within a wide range of link attenuation from Alice to Bob. One can simulate the experimental situation with the model mentioned above. In Fig.\ \ref{figHighDB}, we show the experimentally obtained visibilities and 4-fold coincidence count rate with the black squares and blue circles, respectively. The red and the blue curves are the prediction of our model, which shows good agreement with the experimental data.

\begin{figure}[htbp]
	\centering
	\hspace{0in}
 	\includegraphics[width=4.8in]{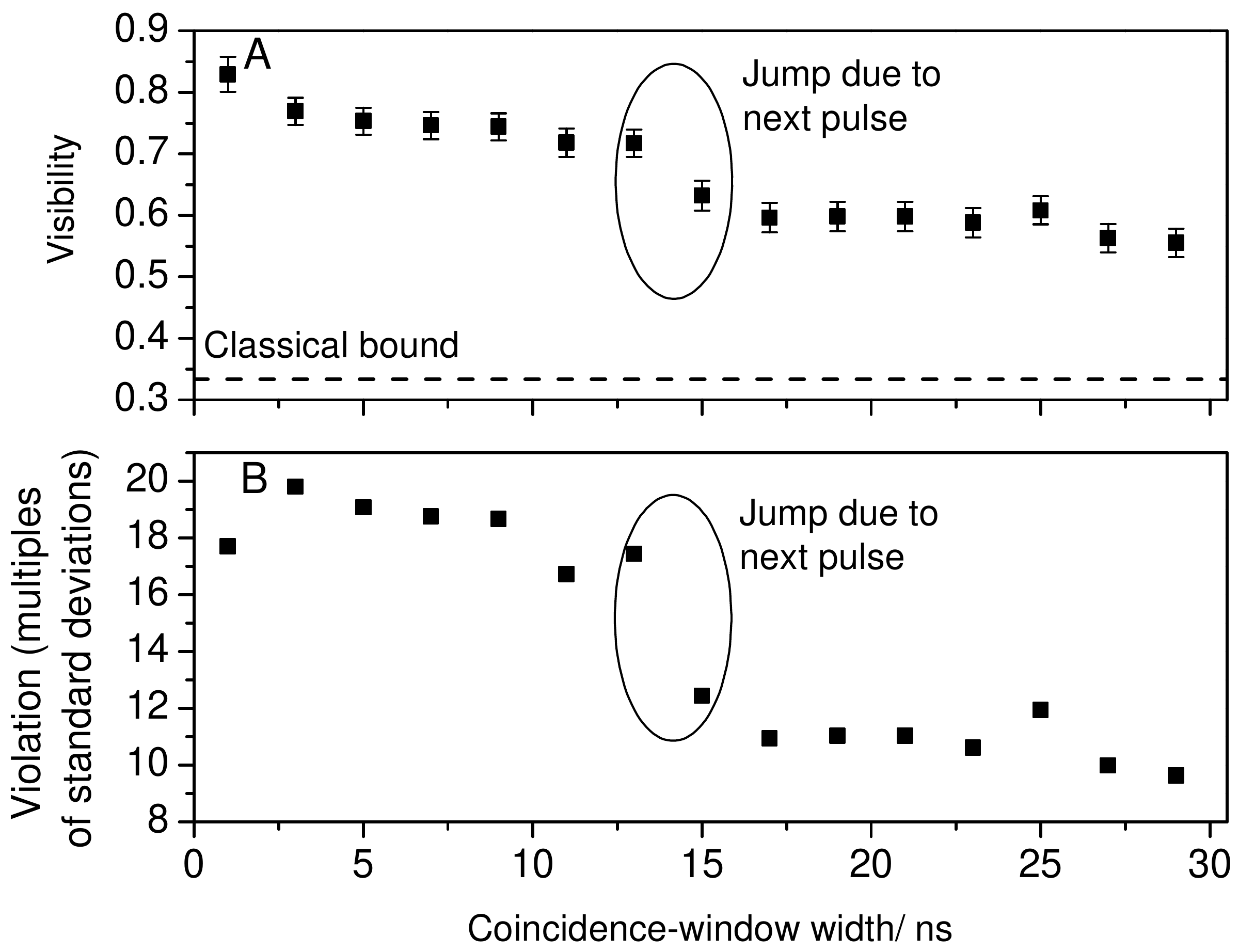}
	\caption{(A) Teleportation visibility vs coincidence-window width. We measure the teleportation visibility with a time-tagging unit and vary the coincidence-window widthin data post processing from 1 ns to 29 ns. The visibility and the numbers of standard deviation (B) violating the classic bound (black straight line in A) of teleportation in general reduces as we increase the coincidence-window width. An obvious drop in both visibility and number of the standard deviation is pointed out. It is due to the increased accidental coincidence counts between consecutive laser pulses, separated by 12.5 ns. This set of data is measured under 31~dB, in 2 hours.}
	\label{VisVsCCW}
\end{figure}

In order to increase the maximum attenuation where teleportation is still possible, the dark count rate and/or the width of the coincidence time window have to be decreased. To reduce the dark counts, first of all the detectors are mounted in a light-isolating box, stopping stray light from the surroundings reaching the detectors. In addition, we have control over the bias voltage of Bob's detectors. Low bias voltage gives lower dark counts, but also reduces the quantum efficiencies of the detectors and hence effectively increases the link attenuation between Alice and Bob. Also, dark-count rates depend on the cooling temperatures of the APDs. We characterized the quantum efficiency (relative to a reference detector) and dark-count rate of Bob's detectors and include that into our model. By cooling the APDs down to $-40$ $^{\circ}$C, the dark count rates of D5 and D6 are about 180 Hz and 400 Hz, respectively. We find the optimal bias voltages and cooling temperatures that give the best SNR or, equivalently, the best teleportation results under a given amount of attenuation. By using alternative APDs~\cite{Stipcevic2010} or stronger cooling means~\cite{Kim2011}, one can further reduce the dark counts and improve the SNR. We did not include the inefficient detectors and loss from Bob's optical elements in the attenuation analysis in any of our experiments. The attenuations we specified include only the attenuations induced by using the neutral density filter and the free-space link.

From Eq.~(\ref{eqSNR}), one can see that another important factor is the temporal length of the coincidence-window width, $\tau$. We use a time-tagging unit to record the arrival time of the BSM results and Bob's detection events relative to an internal clock with 156 ps resolution. Then we process the cross-correlations of these time stamps with different coincidence-window width. As shown in Fig.\ \ref{VisVsCCW}(A) and \ref{VisVsCCW}(B), the visibility and the number of standard deviations violating the classical bound (corresponding to a visibility of 1/3) drop as we increase the size of the coincidence time window, respectively. Note that there is an obvious drop in both visibility and number of the standard deviations due to the increased accidental coincidence counts between consecutive laser pulses, which are separated by 12.5 ns.

In conclusion, we present a high-fidelity quantum teleportation experiment over a high-loss free-space channel. Our work is an important step towards future quantum teleportation experiment between satellites and ground stations. The expected optical link attenuation between a ground based transmitter (sending aperture with 100 cm in diameter) to a low earth orbiting (LEO) satellite receiver (receiving aperture with 13 cm in diameter) is about $-35$~dB~\cite{Aspelmeyer2003b}. It is shown that quantum teleportation is experimentally feasible even under an attenuation as high as $-36$~dB (Fig.\ \ref{figHighDB}), which corresponds to a ground-LEO link attenuation regime.

\section*{Acknowledgment}

The authors acknowledge support from European Space Agency (Contract
4000104180/11/NL/AF), the Austrian Science Foundation (FWF) under project SFB F4008, QTS project (No. 828316) funded by FFG within the ASAP 7 program, the project of the European Commission Q-ESSENCE (No. 248095), ERC Advanced Senior Grant (QIT4QAD), the John Templeton Foundation, as well as SFB-FOQUS and the Doctoral Program CoQuS of the Austrian Science Fund (FWF).


\begin{thebibliography}{99}

\bibitem{Bennett1993} C. H. Bennett, G. Brassard, C. Cr\'{e}peau, R. Jozsa, A. Peres, and W. K. Wootters, ``Teleporting an Unknown Quantum State via Dual Classical and Einstein-Podolsky-Rosen Channels,'' Phys. Rev. Lett. \textbf{70}, 1895--1899 (1993).

\bibitem{Bouwmeester1997} D. Bouwmeester, J.~W. Pan, K. Mattle, H. Weinfurter, and A. Zeilinger, ``Experimental quantum teleportation,'' Nature {\bf 390}, 575--579 (1997).

\bibitem{Boshi1998} D. Boschi, S. Branca, F. De Martini, L. Hardy, and S. Popescu, ``Experimental Realization of Teleporting an Unknown Pure Quantum State via Dual Classical and Einstein-Podolsky-Rosen Channels,'' Phys. Rev. Lett. \textbf{80}, 1121--1125 (1998).

\bibitem{Gottesmann1999}D. Gottesmann and I. L. Chuang, ``Quantum Teleportation is a Universal Computational Primitive,'' Nature \textbf{402}, 390--393 (1999).

\bibitem{Knill2001} E. Knill, R. Laflamme, and G. J. Milburn, ``A scheme for efficient quantum computation with linear optics,'' Nature {\bf 409}, 46--52 (2001).

\bibitem{Ladd2010} T.~D. Ladd, F. Jelezko, R. Laflamme, Y. Nakamura, C. Monroe, and J.~L. O'Brien, ``Quantum computers,'' Nature {\bf 464,} 45--53 (2010).

\bibitem{Yurke1992} B. Yurke and D. Stoler, ``Bell's-inequality experiments using independent-particle sources," Phys. Rev. A \textbf{46}, 2229--2234 (1992).

\bibitem{Zukowski1993} M. \v{Z}ukowski, A. Zeilinger, M. A. Horne, and A. K. Ekert, ```Event-ready-detectors' Bell experiment via entanglement swapping,'' Phys. Rev. Lett. \textbf{71}, 4287--4290 (1993).

\bibitem{Briegel1998} H.~J. Briegel, W. D\"{u}r, J.~I. Cirac, and P. Zoller, `` Quantum Repeaters: The Role of Imperfect Local Operations in Quantum Communication,'' Phys. Rev. Lett. {\bf 81}, 5932--5935 (1998).

\bibitem{Duan2001} L.~M. Duan, M.~D. Lukin, J.~I Cirac, and P. Zoller, ``Long-distance quantum communication with atomic ensembles and linear optics,'' Nature {\bf 414,} 413--418 (2001).

\bibitem{Bose1998} S. Bose, V. Vedral, and P. L. Knight, ``Multiparticle generalization of entanglement swapping,'' Phys. Rev. A \textbf{57}, 822--829 (1998).

\bibitem{Marcikic2003} I. Marcikic, H. De Riedmatten, W. Tittel, H. Zbinden, and N. Gisin, ``Long distance teleportation of qubits at telecommunication wavelengths,'' Nature \textbf{421}, 509--513 (2003).

\bibitem{Ursin2004} R. Ursin, T. Jennewein, M. Aspelmeyer, R. Kaltenbaek, M. Lindenthal, P. Walther, and A. Zeilinger, ``Quantum teleportation across the Danube,'' Nature \textbf{430}, 849 (2004).

\bibitem{Hughes2002} R. J. Hughes, J. E. Nordholt, D. Derkacs, and C.~G. Peterson, ``Practical free-space quantum key distribution over 10 km in daylight and at night,'' New J. Phys. \textbf{4}, 43.1--43.14 (2002).

\bibitem{Kurtsiefer2002} C.~Kurtsiefer, P.~Zarda, M.~Halder, H. Weinfurter, P.~M. Gorman, P.~R. Tapster, and J.~G. Rarity, ``Quantum cryptography: a step towards global key distribution,'' Nature \textbf{419}, 450 (2002).

\bibitem{Aspelmeyer2003} M.~Aspelmeyer, H.~R.~B\"{o}hm, T.~Gyatso, T.~Jennewein, R.~Kaltenbaek, M.~Lindenthal, G.~Molina-Terriza, A.~Poppe, K.~Resch, M.~Taraba, R.~Ursin, P.~Walther, and A.~Zeilinger, ``Long-distance free-space distribution of quantum entanglement,'' Science \textbf{301}, 621--623 (2003).

\bibitem{Resch2005} K.~Resch, M.~Lindenthal, B.~Blauensteiner, H.~R.~B\"{o}hm, A.~Fedrizzi, C.~Kurtsiefer, A.~Poppe, T.~Schmitt-Manderbach, M.~Taraba, R.~Ursin, P.~Walther, H.~Weier, H.~Weinfurter, and A.~Zeilinger, ``Distributing entanglement and single photons through an intra-city free-space quantum channel,'' Opt. Express \textbf{13}, 202--209 (2005)

\bibitem{Peng2005} C.-Z.~Peng, T.~Yang, X.-H.~Bao, J.~Zhang, X.-M.~Jin, F.-Y.~Feng, B.~Yang, J.~Yang, J.~Yin, Q.~Zhang, N.~Li, B.-L.~Tian, and J.-W.~Pan, ``Experimental free-space distribution of entangled photon pairs over a noisy ground atmosphere of 13 km: towards satellite-based global quantum communication,'' Phys. Rev. Lett. \textbf{94}, 150501 (2005).

\bibitem{Schmitt-Manderbach2007} T.~Schmitt-Manderbach, H.~Weier, M.~F\"{u}rst, R.~Ursin, F.~Tiefenbacher, T.~Scheidl, J.~Perdigues, Z.~Sodnik, C.~Kurtsiefer, J.~G.~Rarity, A.~Zeilinger, and H.~Weinfurter, ``Experimental Demonstration of Free-Space Decoy-State Quantum Key Distribution over 144 km,'' Phys. Rev. Lett. \textbf{98}, 010504 (2007).

\bibitem{Ursin2007} R. Ursin, F. Tiefenbacher, T.~Schmitt-Manderbach, H.~Weier, T.~Scheidl, M.~Lindenthal, B.~Blauensteiner, T.~Jennewein, J.~Perdigues, P.~Trojek, B.~\"{O}mer, M.~F\"{u}rst, M.~Meyenburg, J.~Rarity, Z.~Sodnik, C.~Barbieri, H.~Weinfurter, and A.~Zeilinger, ``Free-space distribution of entanglement and single photons over 144 km,'' Nat. Phys. \textbf{3}, 481--486 (2007).

\bibitem{Villoresi2008} P. Villoresi, T.~Jennewein, F.~Tamburini, M.~Aspelmeyer, C.~Bonato, R.~Ursin, C.~Pernechele, V.~Luceri, G.~Bianco, A.~Zeilinger, and C.~Barbieri, ``Experimental verification of the feasibility of a quantum channel between space and Earth,'' New J. Phys. \textbf{10}, 033038 (2008).

\bibitem{Fedrizzi2009} A. Fedrizzi, R.~Ursin, T.~Herbst, M.~Nespoli, R.~Prevedel, T.~Scheidl, F.~Tiefenbacher, T.~Jennewein, and A.~Zeilinger, ``High-fidelity transmission of entanglement over a high-loss free-space channel,'' Nat. Phys. \textbf{5}, 389--392 (2009).

\bibitem{Scheidl2009} T.~Scheidl, R.~Ursin, A.~Fedrizzi, S.~Ramelow, X.-S.~Ma, T.~Herbst, R.~Prevedel, L.~Ratschbacher, J.~Kofler, T.~Jennewein, and A.~Zeilinger, ``Feasibility of 300 km quantum key distribution with entangled states,'' New J. Phys. \textbf{11} 085002 (2009).

\bibitem{Jin2010} X.-M.~Jin, J.-G.~Ren, B.~Yang, Z.-H.~Yi, F.~Zhou, X.-F.~Xu, S.-K.~Wang, D.~Yang, Y.-F.~Hu, S.~Jiang, T.~Yang, H.~Yin, K.~Chen, C.-Z.~Peng, and J.-W.~Pan, ``Experimental free-space quantum teleportation,''  Nat. Photon. \textbf{4}, 376--381 (2010).

\bibitem{Scheidl2010} T.~Scheidl, R.~Ursin, J.~Kofler, S.~Ramelow, X.-S.~Ma, T.~Herbst, L.~Ratschbacher, A.~Fedrizzi, N.~K.~Langford, T.~Jennewein, and A.~Zeilinger ``Violation of local realism with freedom of choice,'' Proc. Natl. Acad. Sci. USA. \textbf{107}, 19709--19713 (2010).

\bibitem{Wootters1982} W.~K. Wootters and W.~H. Zurek ``A single quantum cannot be cloned,'' Nature \textbf{299}, 802--803 (1982).

\bibitem{Calsamiglia2001} J. Calsamiglia and Norbert L\"{u}tkenhaus ``Maximum efficiency of a linear-optical Bell-state analyzer,'' Appl. Phys. B \textbf{72}, 67--71 (2001).

\bibitem{Kwiat1995}
P.~G. Kwiat, K.~Mattle, H.~Weinfurter, A.~Zeilinger, A.~V.~Sergienko, and Y. Shih, ``New High-Intensity Source of Polarization-Entangled Photon Pairs,'' Phys. Rev. Lett. {\bf 75}, 4337--4341 (1995).

\bibitem{White1999} A.~G. White, D. F.~V. James, P.~H. Eberhard, and P.~G. Kwiat, ``Nonmaximally entangled states: Production,
  characterization, and utilization,'' Phys. Rev. Lett. \textbf{83}, 3103--3107 (1999).

\bibitem{James2001} D. F.~V. James, P.~G. Kwiat, W.~J. Munro, and A.~G. White ``Measurement of qubits,'' Phys. Rev. A, \textbf{64}, 052312 (2001).

\bibitem{Popescu1994} S. Popescu, ``Bell's inequalities versus teleportation: What is nonlocality?'' Phys. Rev. Lett. \textbf{72}, 797--799 (1994).

\bibitem{NielsenChuang2000} M. Nielsen and I. Chuang ``\textit{Quantum computation and quantum information},'' 377, 393 (Cambridge Univ. Press, Cambridge, 2000).

\bibitem{Halder2007} M. Halder, A.~Beveratos, N.~Gisin, V.~Scarani, C.~Simon, and H.~Zbinden, ``Entangling independent photons by time measurement,'' Nat. Phys. \textbf{3}, 692--695 (2007).

\bibitem{Stipcevic2010} M. Stip\v{c}evi\'{c}, H. Skenderovi\'{c}, and D. Gracin ``Characterization of a novel avalanche photodiode for single photon detection in VIS-NIR range,'' Opt. Express. \textbf{18}, 17448--17459 (2010).

\bibitem{Kim2011} Y.-S. Kim, Y.-C. Jeong, S. Sauge, V. Makarov, and Y.-H. Kim ``Ultra-low noise single-photon detector based on Si avalanche photodiode,'' Rev. Sci. Instrum. \textbf{82}, 093110 (2011).

\bibitem{Aspelmeyer2003b} M. Aspelmeyer, T. Jennewein, M. Pfennigbauer, W. Leeb, and A. Zeilinger ``Long-distance quantum communication with entangled photons using
satellites,'' IEEE J. Sel. Top. Quant. Electron. \textbf{9}, 1541--1551 (2003).


\end{thebibliography}
\end{document}